\documentclass[prl,twocolumn,superscriptaddress,floatfix,amsmath,nofootinbib,amssymb]{revtex4}

\usepackage{graphicx}% Include figure files
\usepackage{dcolumn}% Align table columns on decimal point
\usepackage{mathrsfs}
\usepackage{bm}
\usepackage{color,xspace}
\usepackage[small,compact,raggedright]{titlesec}
\usepackage{epstopdf}

\newcommand{\ket}[1]{\left| {#1} \right\rangle}
\newcommand{\bra}[1]{\left\langle {#1} \right|}
\newcommand{\ematriz}[3]{\left\langle {#1} \left|{#2}\right|{#3}\right\rangle}	

\newcommand{\proj}[2]{\left| {#1} \right\rangle\!\left\langle {#2} \right|}

\newcommand{\tr}{\operatorname{tr}}

\begin{document}

\title{Using Berry's phase to detect the Unruh effect at lower accelerations}
\author{Eduardo Mart\'{i}n-Mart\'{i}nez}
\affiliation{Instituto de F\'{i}sica Fundamental, CSIC, Serrano 113-B, 28006 Madrid, Spain}
\author{Ivette Fuentes}
\affiliation{School of Mathematical Sciences, University of Nottingham, Nottingham NG7 2RD, United Kingdom}
\author{Robert B. Mann}
\affiliation{Department of Physics \& Astronomy, University of Waterloo, Waterloo, Ontario Canada N2L 3G1}

\begin{abstract}
We show that a detector acquires a Berry phase due to its motion in spacetime. The phase is different in the inertial and accelerated case as a direct consequence of the Unruh effect. We exploit this fact to design a novel method to measure the Unruh effect.  Surprisingly, the effect is detectable for accelerations $10^9$ times smaller than previous proposals sustained only for times of nanoseconds.
\end{abstract}

\date{December 10th, 2010}


%\keywords{Entanglement, gravitational collapse, black holes, Hawking radiation}

\maketitle

 In the Unruh effect \cite{Unruh0,bigreview} the vacuum state of a quantum field corresponds to a thermal state when described by uniformly accelerated observers. Its 
direct detection  is unfeasible with current technology since the Unruh temperature is smaller than 1 Kelvin even for accelerations
as high as $10^{21}$ $\text{m}/\text{s}^2$.  Sustained accelerations higher than $10^{26}$ $\text{m}/\text{s}^2$ are required to detect the effect \cite{ChenTaj,bigreview}. In this letter we show that the state of a moving detector coupled to the field acquires a Berry phase \cite{Berryoriginal} due to its movement in spacetime. This geometric phase, which is a function of the detector's trajectory, encodes information about the Unruh temperature and it is observable for accelerations as low as $10^{17}$ $\text{m}/\text{s}^2$. Such acceleration must be sustained only for a few nanoseconds. Our results enormously simplify the challenge of measuring the Unruh effect with present technology since producing extremely high accelerations and measuring low temperatures were the main obstacles involved in its detection.  The results presented here are independent of specific experimental implementations; however, we propose a possible scheme for the detection of this phase.

Finding indisputable corroboration of the Unruh effect is one of the main experimental goals of our time \cite{experiments,bigreview}. The effect is one of the best known predictions  of  quantum field theory incorporating general relativity. However, its  very existence has been subject to lengthy controversy  \cite{sceptic}.  Its observation would provide not only an end to such discussion but also experimental support for  Hawking radiation and black hole evaporation, given the deep connection between these phenomena \cite{Hawking}.  Detection of the Unruh effect would have an immediate impact in many fields such as astrophysics \cite{Astronature}, cosmology \cite{Cosmo}, black hole physics \cite{Bholes}, particle physicsÊ\cite{Base},  quantum gravity \cite{Qg} and relativistic quantum information \cite{Alicefalls}. 

Efforts toward finding evidence of the Unruh and Hawking effects also include proposals in analog systems such as fluids \cite{Unruhan}, Bose-Einstein condensates \cite{garay}, optical fibers \cite{optfib}, slow light \cite{slowlight},  superconducting circuits \cite{supercond} and trapped ions \cite{ions}.  Even in such systems,  analog effects produce temperatures of the order of nanokelvin that remain difficult to detect.

Interestingly, it has gone unnoticed that Berry's phase can be employed to detect the Unruh effect. Berry showed that an eigenstate of a quantum system acquires a phase, in addition to the usual dynamical phase, when the parameters of its Hamiltonian are varied in a cyclic and adiabatic fashion \cite{Berryoriginal}.  In the case of a point-like detector interacting with a quantum field, the movement of the detector in spacetime produces, under certain conditions, the cyclic and adiabatic evolution that gives rise to Berry's phase. We will show that the Berry phase for an inertial detector
differs from that of an accelerated one.  This difference arises due to the Unruh effect: one detector interacts with the vacuum state, the other with a thermal state. The Berry phase of an accelerated detector depends on the Unruh temperature. Surprisingly, we find that this phase is observable for detectors moving with relatively low accelerations, making the detection of the Unruh effect  accessible with current technology.
\begin{figure}[h]
\begin{center}
\includegraphics[width=.36\textwidth]{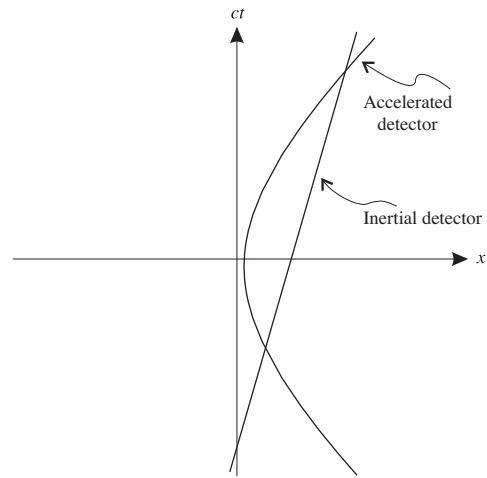}
\caption{Trajectories for an inertial and accelerated detector.}
\label{setup}
\end{center}
\end{figure}

In our analysis, we consider a massless scalar field in the vacuum state from the perspective of inertial observers moving in a flat ($1+1$)-dimensional spacetime. The same state of the field from the perspective of uniformly accelerated observers corresponds to a thermal state whose temperature  is the so-called Unruh temperature ${T_U=\hbar a / (2\pi c k_\text{B})}$ where $a$ is the observer's acceleration, $c$ the speed of light and $k_\text{B}$ Boltzmann's constant.

In order to show evidence of this effect, we consider a point-like detector endowed with an internal structure that couples linearly to the scalar field $\phi(x(t))$ at a point $x(t)$ corresponding to the world line of the detector.  When the detector is considered to be a harmonic oscillator with ladder operators $b^{\dagger}$ and $b$, the interaction Hamiltonian is given by $H_I\propto (b^\dagger + b)\hat \phi(x(t))$ where $(b^\dagger + b)$ is the detector's position operator. This model is a type of Unruh-DeWitt detector \cite{bigreview} which has been previously studied in \cite{LinHu}.  
In a realistic scenario the oscillator couples to a peaked distribution of field modes. However if the distribution can be contrived to approach a delta function we can assume that only one mode of the field is coupled to the detector. In this case the field operator takes the form $\hat\phi(x(t))\approx\hat\phi_k(x(t))\propto \left[a\, e^{i(kx-\Omega_a t)}+a^\dagger\, e^{-i(kx-\Omega_a t)}\right]$, where $a^{\dagger}$ and $a$ are creation and anihilation operators associated to the field mode $k$ with frequency $|k|=\Omega_a$. The Hamiltonian is therefore
\begin{equation}\label{goodham2}
 H_T\!=\!\Omega_a a^\dagger a+\Omega_b b^\dagger b + \lambda (b+b^\dagger)[a^\dagger e^{i(kx-\Omega_a t)}+ a e^{-i(kx-\Omega_a t)}]
 \end{equation}
 where $\Omega_a$ and $\Omega_b$ are the field and atom frequencies respectively and $\lambda$ is the coupling frequency. 
 
The single mode interaction can be engineered, for instance, by employing a cavity. Considering that the cavity field modes have very different frequencies and one of them is close to the detector's natural frequency,   the  detector effectively interacts only with this single mode. It is well known that introducing a cavity is problematic since the boundary conditions may inhibit the Unruh effect. However, this problem is solved by allowing the cavity to be transparent to the field mode the detector couples to.  Therefore this single mode is a global mode.  In a realistic situation, the cavity would be transparent to a frequency window which is experimentally controllable. It is then an experimental task to reduce the window's width as required. 

 Although calculations involving  Unruh-DeWitt detectors usually employ interaction or Heisenberg pictures (as transition probabilities are more conveniently calculated), in \eqref{goodham2} we employ a mixed picture where the detector's operators are time independent. This situation is mathematically more convenient for Berry phase calculations; the results are,  as expected, picture independent. 

The Hamiltonian \eqref{goodham2} can be diagonalized analytically; its eigenstates are $U^{\dagger}|N_a N_b\rangle$, where $|N_a N_b\rangle$ are eigenstates of $H_0(\omega_a,\omega_b)=\omega_a\, a^\dagger a+\omega_b\, b^\dagger b$ and $U=S_a S_b  D_{ab}\hat S_b R_a$. The operators
\begin{equation}
\begin{array}{ll}
\nonumber D_{ab}=\exp\big[s (a^\dagger b -  a b^\dagger)\big],&S_{a}=\exp\big[\frac12 u({a^\dagger}^2 - a^2)\big],\\*
\nonumber S_{b}=\exp\big[\frac12 v(b^2 - {b^\dagger}^2) \big],&\hat S_{b}=\exp\big[p\, ({b^\dagger}^2 - b^2)\big]
\end{array}
\end{equation}
 and $R_a=\exp\big(-i\varphi\, {a^\dagger a}\big)$ are the two-mode displacement, single-mode squeezing and phase rotation operators \cite{Scullybook}, respectively. 

The parameters $u,v,s,p,\omega_a,\omega_b$ are functions of $\lambda$ and the detector frequencies $\Omega_a,\Omega_b$. Their functional form is obtained when diagonalizing $H_T$. Only three parameters turn out to be independent, and we can write  $u,s$ and  $p$  in terms of $v, \omega_a$ and $\omega_b$.  Details will appear in a forthcoming paper \cite{IER}.

The phase of the field operators $\varphi=kx-\Omega_a t$, where  we have used Minkowski coordinates $(t,x)$
(a convenient choice for inertial observers),  varies  due to the time evolution along the detector's trajectory.  Therefore, the displacement of the detector in spacetime generates a cyclic change in the Hamiltonian.  The parameter $\varphi$ completes a $2\pi$ cycle in a period of time $\Delta t\sim\Omega_a^{-1}$.

Consider a scenario such that before the interaction between the field and the detector is switched on, the field is in the vacuum state and the detector in the ground state $\ket{0_f0_d}$. Employing the sudden approximation, we find that after the coupling is suddenly switched on the state of the system is
\begin{equation}\label{adiab}
\ket{\psi_{00}}=\sum_{n,m} \ematriz{n_fm_d}{U}{00}U^\dagger \ket{n_fm_d}
\end{equation}
In the coupling regimes we consider, the probability of detector excitation due to the sudden switching on is negligible. For small $\lambda$ and for any value of $\Omega_a$ and $\Omega_b$, the state  $\ket{0_f0_d}$ is an approximate eigenstate of the operator $S_aS_bD_{ab}\hat S_b R_a$ with eigenvalue 1. Therefore, under these conditions the state of the system  immediately after the interaction has been switched on is  $U^\dagger\ket{0_f0_d}$.

Now we investigate under what conditions the time evolution of the coupled field-detector system is adiabatic.  During the evolution the ground state $U^\dagger\ket{0_f0_d}$ does not become degenerate and the energy gap between the ground and first excited state is time-independent. For small but realistic values of $\lambda$, energy conservation ensures a negligible probability for the system to evolve into an excited state (an explicit calculation of the probability of excitation is given in \cite{Ivy}).  In this case, the evolution due to the movement of the detector in spacetime  is adiabatic since the ground state of the Hamiltonian $H(t_0)$ evolves after a time  $t-t_0$ to the ground state of the Hamiltonian $H(t)$. 

After finding under which conditions the evolution is cyclic and adiabatic we are able to compute the Berry phase
$\gamma$ acquired by the state $U^\dagger\ket{0_f0_d}$ 
after a cycle in $\varphi$. For an eigenstate $\ket{\psi(t)}$ of $H_T$, $i\gamma=\oint_R\, \bm A \cdot  \text{d}\bm R$
where $A_i=\bra{\psi(t)}\partial_{R_i}\ket{\psi(t)}$ 
 and $R$ is a closed trajectory in the parameter space $\{R_1(t),\dots,R_k(t)\}$ on which $H_T$ depends \cite{Berryoriginal} . 
 For our particular case of the inertial detector in our scenario, we obtain
\begin{align}
\nonumber \frac{\gamma_{I}}{2\pi}=&\,\frac{\omega_a \sin^2 v \sinh [2(C-v)]+\omega_b\,\sinh(2v)\sinh^2 (C-v)}{\omega_a\, \sinh[2(C-v)]+\omega_b\,\sinh (2v)},
\end{align}
where $C=\frac{1}{2}\ln\left(\omega_a/\omega_b\right)$ with  $\omega_a/\omega_b>e^{2v}$. Here the label $I$ denotes that the phase corresponds to the inertial detector. Note that the phase is identical for all inertial trajectories.  In what follows, we  show that, as a direct consequence of the Unruh effect, the phase is different for accelerated detectors. 

Computing the Berry phase in the accelerated case is slightly more involved.  A convenient choice of coordinates for the accelerated detector are Rindler coordinates $(\tau,\xi)$. In this case $\varphi\!=\!|\Omega_a|\xi-\Omega_a\tau$. The evolution is cyclic after a time $\Delta \tau = \Omega_a^{-1}$.  Adiabaticity can also be ensured  in this case since the probability of excitation is negligible for the accelerations we will later consider \cite{Scully,bigreview}.  

We assume that identical detectors couple  to the field in both inertial and accelerated cases. Hence they couple to the same proper frequency (the frequency in the reference frame of the detector). Note that these frequencies are not the same from the perspective of any inertial observer.  Although $H_T$ in \eqref{goodham2} has the same form in both scenarios, in the inertial case $a,a^\dagger$ are Minkowski operators, whereas for the accelerated detector  they correspond to Rindler operators. To make this distinction clear, from now on we denote $U^\dagger_\text{M}$ and $U^\dagger_\text{R}$ with the understanding that the operators involved are Minkowski and Rindler, respectively.  For accelerated observers  the state of the field is not pure but mixed, a key distinction from  the inertial case.   Expressing the state of the field and  detector  in the basis of an accelerated observer, the state $\proj{0_f}{0_f}$ transforms  to  the thermal Unruh state $\rho_f$ \cite{Unruh0,Alicefalls}. 
Therefore, before turning on the interaction between the field and the detector, the system is in the mixed state  $\rho_f\otimes{\proj{0_d}{0_d}}$. When the interaction is suddenly switched on, a general state $\ket{N_f 0_d}$ evolves, in our coupling regime, very close to a superposition of eigenstates  $U_R^\dagger\ket{i_f j_d}$ where $N_f=i_f+j_d$. If immediately after switching on the interaction we verify that the detector is still in its ground state (by making a projective measurement) we can assure that the state of the joint system is  $\rho_T= U_R^\dagger \left(\rho_f\otimes{\proj{0_d}{0_d}}\right)U_R$.

Calculating the mixed state Berry phase \cite{Vlatko} we find
\[\gamma_a=\gamma_{I}-\operatorname{Arg}\left(\cosh^2 q - e^{2\pi\,i G } \sinh^2 q\right)\]
where
$\gamma_{I}$ is the inertial Berry phase, $q=\arctan\big(e^{-\pi\Omega_a c/a}\big)$ and 
\[G= \frac{\omega_b\,  \sinh(2v)\cosh[2(C-v)]}{\omega_a\sinh[2(C-v)]+\omega_b\sinh(2v)}\]
 depends  on the detector parameters.
 
We now compare the Berry phase acquired by the detector in the inertial and accelerated cases.
After a complete cycle in the parameter space (with a proper time $\Omega_a^{-1}$) the phase difference between an inertial and an accelerated detector is
$\delta =\gamma_{I}-  \gamma_a$.  

\begin{figure}
\begin{center}
\includegraphics[width=.42\textwidth]{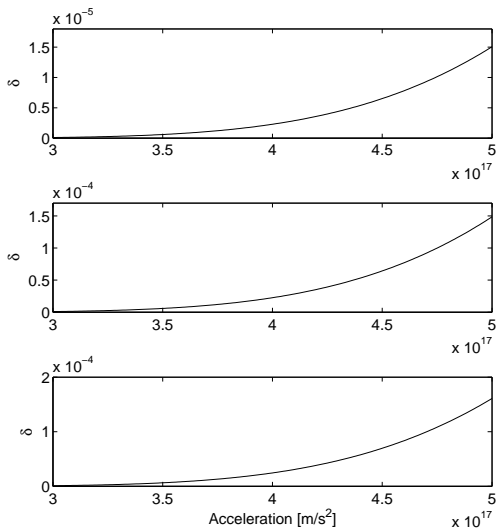}
\caption{$\delta$ for each cycle as a function of the acceleration for three different scenarios. First scenario (top): $\Omega_a\simeq 2.0$ GHz  $\Omega_b\simeq 2.0$ GHz $\lambda\simeq$ 34  Hz.
Second scenario (middle): $\Omega_a\simeq 2.0$ GHz  $\Omega_b\simeq 2.0$ GHz  $\lambda\simeq$  0.10  KHz.
Third scenario (bottom): $\Omega_a\simeq 2.0$ GHz  $\Omega_b\simeq 2.0$ GHz  $\lambda\simeq$ 0.25  KHz.}
\label{deph}
\end{center}
\end{figure}

In figure \ref{deph} we plot the phase difference $\delta$ as a function of the acceleration corresponding to choosing physically relevant frequencies of atom transitions \cite{revat,Scully} coupled to the electromagnetic field (in resonance with the field mode they are coupled to) for the microwave regime ($2.0$ GHz) and for three different coupling strengths: 1) $\lambda\simeq$ $34$  Hz,  2) $\lambda\simeq$  $0.10$   KHz,  3)   $\lambda\simeq$ $0.25$  KHz. 

The third case, where the coupling frequency $\lambda \simeq 10^{-7}\,\Omega_a$, corresponds to typical values for atoms in  free space with dipolar coupling  to the field \cite{revat}. For a single cycle (after 3.1 ns) the phase difference is large enough to be detected. The visibility of the interference pattern is given by $V=\sqrt{\tr\left[\proj{0_f0_d}{0_f0_d}(\rho_f\otimes{\proj{0_d}{0_d}})\right]}=\cosh^{-1}q\simeq 1$. Note that the visibility is approximately  unity in all the situations we consider due to the relatively low accelerations involved. 

Since the Berry phase accumulates, we can enhance the phase difference by evolving the system through more cycles. By allowing the system to evolve for the right amount of time, it is possible to produce a maximal phase difference of $\delta=\pi$ (destructive intereference). For example, considering an acceleration  of $a\approx 4.5\cdot10^{17}$ $\text{m}/\text{s}^2$  a maximal phase difference would be produced after $30000$ cycles.  Therefore, given the frequencies considered in our examples, one must allow the system to evolve for 95 $\mu$s. 

Note that for an acceleration of $a\approx 10^{17}$ $\text{m}/\text{s}^2$ the atom reaches speeds of $\approx 0.15c$ after a time $t\approx \Omega_a^{-1}$. 
The longer we allow the system to evolve in order to obtain a larger phase difference, the more relativistic the atom becomes. Therefore,  depending on the particular experimental implementation considered to measure the effect, a compromise between the desired phase difference and feasibility of handling relativistic atoms must be considered.  This experimental difficulty can be overcome by means of different techniques. For example,  since the phase accumulates independently of the sign of the acceleration, one could consider alternating periods of positive and negative acceleration in order to reduce the final speed reached by the atom.  This will also help to cancel the dynamical phase difference between the paths in a specific setting as discussed later. 
%\begin{figure}[h]
%\begin{center}
%\includegraphics[width=.50\textwidth]{vis}
%\caption{Visibility factor as a function of the acceleration for three different scenarios. First scenario (top): $\Omega_a\simeq 0.10$ THz,  $\Omega_b\simeq 0.16$ THz, $\lambda\simeq 1.5\!\cdot\! 10^{-2}$ THz.
%Second scenario (middle): $\Omega_a\simeq 0.10$ THz  $\Omega_b\simeq 0.12$ THz  $\lambda\simeq  5.5\cdot 10^{-3}$ THz.
%Third scenario (bottom): $\Omega_a\simeq 0.10$ THz  $\Omega_b\simeq 0.10$ THz  $\lambda\simeq 2.5\cdot 10^{-3}$ THz.}
%\label{realf}
%\end{center}
%\end{figure}
The Berry phase is always a global phase. In order to detect it, it is necessary to prepare an interferometric experiment.  For example, a detector in a superposition of an inertial and accelerated trajectory would allow for detection of the phase. Any experimental set-up in which such a superposition can be implemented would serve our purposes. A possible scenario can be found in the context of atomic interferometry. This technology has already been successfully employed to measure with great precision general relativistic effects such as time dilation due to Earth's gravitational field \cite{atomGR}.

Consider the detector to be an atom which is introduced into an atomic interferometer after being prepared in its ground state. In one arm of the interferometer we let the atom move inertially. In the other arm we consider a mechanism  which produces a uniform acceleration of the atom. Such mechanism could consist of laser pulses that are prolonged for fractions of nanoseconds.  Laser technology  producing such high accelerations is already available \cite{laser}.  In order for the detector to survive at least long enough to conclude the interference experiment,  the laser pulses must be engineered to create the deep potential wells necessary to accelerate the atom without exciting it.  As long as the atom does not collide with other atoms this seems feasible \cite{ruso}.  An alternative to this is to consider ions or atomic nuclei as detectors which can be accelerated by applying a potential difference in one arm of the interferometer.   While such set-ups are obviously not exempt from technical difficulties,  the experimental challenges involved are expected to be solvable with  near-future technology.  

 Paths (of slightly different length) can be chosen such that the dynamical relative phase cancels. It is sufficient that the dynamical phase difference through both trajectories be equal or smaller than the geometric phase to allow for its detection. Although such cancellation depends upon the specific experimental setup, we find that even for a simple setting with current length metrology technology \cite{Laserscale}, we can control the relative dynamical phase with a precision $\Delta \phi \approx 10^{-8}$, several orders of magnitude smaller than the Berry phase acquired in one cycle. 

Here we have shown that the Unruh effect leaves its footprint in the geometric phase acquired by the the joint state of the field and the detector for time scales of about $5\times10^{-10}$ s. The effect is observable for accelerations as low as $10^{17}$ $\text{m}/\text{s}^2$ and can be maximally enhanced allowing the system to evolve a few microseconds. 

Our theoretical setting is general and independent of any particular implementation, paving the way for future experimental proposals. For instance, by considering detector frequencies in the MHz regime,  the method would allow detection of the Unruh effect for accelerations as low as $10^{14}$ $\text{m}/\text{s}^2$ . For this, other multilevel harmonic systems could be employed as detectors, such as fine structure transitions where frequencies are closer to MHz regime.  Possible experimental implementations of this method are expected to be suggested elsewhere \cite{IER}.

We would like to thank J. Louko, D. Gelbwaser, C. Henkel, D. R\" atzel, A. Dragan, T. Fernholz, L. Hackermuller and P. Kr\"uger for interesting discussions.  I.~F. thanks EPSRC  [CAF Grant EP/G00496X/2] for financial support. E. M-M was supported by a CSIC JAE-PRE Grant scheme, by the Spanish MICINN Project FIS2008-05705/FIS and the QUITEMAD consortium. R.B.M was supported in part by  NSERC.

\end{document}